\documentstyle[12pt, amssymb]{article}
\hoffset =  - 1cm
\topmargin = - 1cm
\begin{document}
\large
\begin{center}
{\bf The probabilitiy distribution density
of random values of squared functional on
Wiener process trajectories}
\medskip

Virchenko Yu.P., Vitokhina N.N.
\end{center}

{\bf 1.}\ We consider the problem about
the distribution density calculation
of the random variable $J_T [w]$ in this work. Here
$$J_T [x]\ =\ \int\limits_0^T x^2 (t) dt \eqno (1)$$
is the functional in the ${\bf L}_2 ([0, T])$ space and
$\{w(t); t \in [0, T]\}$, $T > 0$ are trajectories of
the standard Wiener process on $[0, T]$, i.e. ${\sf E}w^2(t) = t$.
This problem is classical as well as analogous problems
concerns the probability
distribution density  calculation of random variables performed by
additive squared functionals on trajectories of gaussian random processes.
For such processes, the calculation problem connected with characteristic
functions of random values under consideration is solved principally.
Up to present time, a great number specific problems are solved
on the basis of this method which  have
various applications (see \cite{Maz}, \cite{Arato}).
For example, such a result concerns the normal markovian process
(the Ornstein-Uhlenbeck process) has been still obtained
in the work \cite{Zie}.
However, the problem of restoration of the distribution density on the
basis of obtained characteristic functions remains
weakly investigated in the sense of the  constructing of
approximations
with the guaranteed accuracy being suitable for use whereas in the
area of the random variable changing.
The usual approach to the solution of this problem (see, for example,
\cite{Zol}) results in some approximated formulas for distribution densities
suitable for the estimation of great fluctuation probabilities, i.e.
in the asymptotic area $x \to \infty$ of the
random value changing.
Here, we study the calculation problem of successive
approximations of the probability distribution density $f(x)$ connected with
random  values of the functional (1) defined
in any compact interval  $[0, M]$, $M > 0$.

{\bf 2.} Since trajectories $\{w(t); t \ge 0\}$
of the standard Wiener process  ${\sf E}w^2(t) = t$
are continuous with the probability one, then
the random value $J_T[w]$ is determined almost sure for each of them.

The generating function of the random value (1) is evaluated by
the formula (see, for example, \cite{Maz})
$$Q_T(\lambda)\ =\ {\sf E}\exp(- \lambda J_T [w])\ =\
\left[{\rm ch}\left(\lambda^{1/2}T\right)\right]^{-1/2}\,, \quad
Q_T(\lambda) = Q_1 (\lambda T^2)\,.$$
The distribution density $f(x)$ of the random value  $J_T [w]$
is defined by the inverse Laplace transformation
$$f(x)\ =\ \frac 1{2\pi i}\int\limits^{i\infty+c}_{-i\infty+c} e^{\lambda x}
\left[{\rm ch}\left(\lambda^{1/2}T\right)\right]^{-1/2} d\lambda\,, \eqno (2)$$
where $c>$0 and, for the integrated function, the cross-cut
in the complex plane $\lambda$ is done along
the negative part of the real axe.

It is convenient to introduce the density  $g(x) = T^{2}f(T^2 x)$.
The replacement of the integration variable in Eq.(2) gives
$$g(x)\ =\ \frac 1{2 \pi i}\int\limits^{i\infty+c}_{-i\infty+c}
 e^{\lambda x} \left[{\rm ch}\left(\lambda^{1/2}\right)\right]^{-1/2} d\lambda
\ =$$
$$=\ \frac 1{\sqrt 2 \pi i}\int\limits^{i\infty+c}_{-i\infty+c}
\left(\frac {\exp(2\lambda  x -\lambda^{1/2})} {1+\exp(-2\lambda^{-1/2})}
\right)^{1/2} d\lambda \eqno (3)$$
in this case. Let us prove the following theorem.

T h e o r e m.\
{\sl The density $g(x)$ is represented by the following absolutely
converging series
$$g(x)\ =\ \sqrt{\frac 2{\pi x^3}}\,
\sum_{l=0}^\infty (-1)^l\,\frac{(2l)!}
{4^{l}(l!)^2}\,(l+1/4) \exp\left(-\frac {(l+1/4)^2}x \right)\,, \eqno (4)$$
the $N$th remainder of this series is estimated by the value}
$$|g(x)-g_{N-1}(x)| \le \sqrt{\frac 2 {\pi x^3}}\left(
\frac{(2N)!}{4^{N}(N!)^2}\right)(N+1/4) \exp
\left(-\frac {(N+1/4)^2}x \right)\,. \eqno (5)$$

$\square$\ We shall put $c = 0$ in Eq.(3) since singularities of
integrated function are on the negative part of the real axe.
We deform the integration contour to the contour $C$ consisting of the
consecutive transitions of following ways
$\{s-i\varepsilon; s \in (-\infty; 0]\}$, $\{\varepsilon e^{is};
s \in [-\pi/2; \pi/2]\}$, $\{-s +i\varepsilon; s \in [0;+\infty)\}$.
Such a deformation is permissible since
$$\left|{\rm ch}\left(\lambda^{1/2}\right)\right|^2 =
{\rm ch}\left(\lambda^{1/2}\right){\rm ch}\left(
(\lambda^*)^{1/2}\right) =\frac 12\left[{\rm ch}\left(2{\rm Re} (\lambda^{1/2})
\right) + {\rm ch} \left(2i \rm{Im}(\lambda^{1/2})\right)\right]\ >$$
$$>\frac 12\left({\rm ch}\left(2R^{1/2} \cos(\varphi/2)\right)-1\right)
= {\rm sh}^2 \left(R^{1/2}\cos (\varphi/2)\right)\,,$$
where $\lambda = Re^{i\varphi}$
and, on the arch $\{\lambda; \varphi \in [\pi/2;\pi)\}$ of the circle,
the following estimation of the integrated expression module in Eq.(3)
is valid
$$\left| \frac{\exp(\lambda x)} {\left({\rm ch}
(\lambda^{1/2})\right)^{1/2}} \right| \le \frac
{\exp(xR\cos\varphi)} {\left[{\rm sh}(R^{1/2}\cos
(\varphi/2))\right]^{1/2}}\,.$$
It guarantees the fulfillment of the Jordan condition  at $x>0$ on the arch
$R^{1/2} \cos(\varphi/2) < \varepsilon$
at any small $\varepsilon > 0$ since
$\cos\,\varphi <0$. The same takes place for the arch
$\{\lambda; \varphi \in (-\pi, \pi/2]\}$.

In Eq.(3) we shall realize the replacement of the integration variable
$\lambda^{1/2} = q$,
then $\lambda = q^2$, $d\lambda = 2qdq$.
Thus, the contour $C$ in the complex plane $\lambda$ will be transformed
to the line  and, after the transition to the limit  $\varepsilon \to 0$,
it will be transformed to the line  $\{q=is; s \in {\Bbb R}\}$ in the
complex plane $q$.
After these transformations, we have
$$g(x) = \frac{\sqrt2}{\pi i} \int\limits^{i\infty}_{-i\infty}q
\left(\frac {\exp(2q^2 x-q)} {1 + \exp(-2q)} \right)^{1/2} dq\,.$$
Now, we pass to the integration on the variable  $s, q=c+is, dq=ids $.
Then we obtain
$$g(x) = i\frac{\sqrt 2} {\pi} \int\limits^{+\infty}_{-\infty}s
\left(\frac {\exp(-2xs^2 -is)} {1+ \exp(-2is)}\right)^{1/2} ds\,.\eqno (6)$$
In the last integral, we shall produce the shift $s+i(4x)^{-1} \Rightarrow s$
of the integration variable. Therefore, we obtain
$$g(x)\ =\ i\frac{\sqrt{2}} {\pi}
\int\limits^{+\infty}_{-\infty}s\
\left(\frac {\exp(-2x(s -i(4x)^{-1})^2 - (8x)^{-1})} {1+
\exp(-2i(s + i(4x)^{-1}) - (2x)^{-1})} \right)^{1/2} ds = $$
$$=\ \frac{\sqrt{2}} {\pi} \exp(-(16x)^{-1}) \int\limits^{+\infty}
_{-\infty} (is+(4x)^{-1}) \left(\frac {\exp(-2xs^2)} {1+ e^{-1/(2x)}
\exp(-2is)} \right)^{1/2} ds\,. \eqno (7)$$

We decompose the denominator of the integrated expression in Eq.(7)
into the series converged at any $x>0$ and at any  $s \in {\Bbb R}$
$$(1+ e^{-1/(2x)} \exp(-2is))^{-1/2} = \sum_{l=0}^\infty
(-1)^l \frac{(2l-1)!!} {2^l l!} \exp(-l/(2x))\exp(-2ils).$$
The convergence is uniform in any area $[0, M] \times
{\Bbb R}$ in the plane $(x, s)$, $M > 0$.
Substituting the last expression in Eq.(7), we obtain
$$g(x) = \frac{\sqrt{2}} {\pi} \exp(-(16x)^{-1}) \int\limits^{+\infty}
_{-\infty} (is+ (4x)^{-1}) \exp(-xs^2)\ \times \phantom {AAAAAAAAAAA}$$
$$\phantom {AAAAAAAAA}\times\ \left[\,\sum_{l=0}^\infty
(-1)^l \frac{(2l)!} {4^l (l!)^2} \exp(-2ils) \exp(-l/ (2x))\right]ds =$$
$$= \frac{\sqrt{2}} {\pi} \exp(-(16x)^{-1})\left[\sum_{l=0}^\infty
(-1)^l \frac{(2l)!} {4^l (l!)^2} \int\limits^{+\infty}_{-\infty}
(is+ (4x)^{-1})
\exp(-xs^2)\ \times \phantom{AAAAA} \right.$$
$$\left.
\phantom {AAAAAAAAAAAAA}\times\ \exp\left( -x (s+il/x)^2 - x^{-1}l(l + 1/2)
\right)\right]ds\,.$$

The transposition of summation and integration is based on the
uniform convergence of the series on $s$ at any fixed $x$.

In each summand of the sum, we shall produce the shift
$s+il/x \Rightarrow s$ of the integration variable,
$$g(x) = \frac{\sqrt2} {\pi} \exp(-(16x)^{-1})\times\
\phantom{AAAAAAAAAAAAAAAAA}$$
$$\phantom {AA}\times \int\limits^{+\infty}_{-\infty}\left[\,
\sum_{l=0}^{\infty} (-1)^l \frac{(2l)!} {4^l (l!)^2}
\left(is+ x^{-1}\left[l + 1/4\right]\right)
\exp\left(-xs^2-x^{-1}l(l + 1/2) \right)\right]ds\,.\eqno (8)$$
Let us represent the integral as the sum of two integrals in accordance with
the expression in the bracket before exponent.
The integral corresponding the summand $is$ converts to the zero
$$i\int\limits^{+\infty}_{-\infty} s
\left[\,\sum_{l=0}^\infty (-1)^l
 \frac{(2l)!} {4^l (l!)^2} \exp(-xs^2)\exp\left(-x^{-1}l(l + 1/2) \right)
\right]ds = 0\,,$$
due to the oddness of the integrand function.
The integral corresponding the summand $x^{-1}\left[l + 1/4\right]$,
transforms as follows
$$x^{-1}\left[\int\limits^{+\infty}_{-\infty}
\exp(-xs^2) ds\right]\left[\,\sum_{l=0}^\infty
(-1)^l \frac{(2l)!} {4^l (l!)^2} \left[l+1/4\right]
\exp\left(-x^{-1}l(l +1/2) \right)\right]\,.$$
Substitution of this expression in Eq.(8) taking into account
the value of the Poisson integral results in the formula (4).

Since the series (4) is alternating in sign
then the remainder of the series does
not exceed the first summand among rejected ones.
Hence, the estimation (5) is valid. \ $\blacksquare$

C o r o l l a r y.\
{\sl The following estimation takes place}
$$|g(x)-g_{N-1}(x)|\ <\ \frac 3{2e^2}N^{-5/2}\,.\eqno (9)$$
$\square$\
Let us estimate the remainder of the series (4).
For this purpose, on the basis Eq.(4), we write down
the density $g(x)$ in the form
$$g(x)\ =\ \sqrt{\frac 2{\pi}}\,\sum_{l=0}^\infty (-1)^l\,
a_l h_l (x)\,, $$
where
$$a_l = \frac{(2l)!}{4^{l}(l!)^2}\,(l+1/4)\,, \quad h_l (x) =
x^{- 3/2}\,\exp\left(-\frac {(l+1/4)^2}x \right),.$$
Also, we may find maximums on $x$ of the functions $h_N (x)$, $n =1, 2, 3, ...
$\,.
Equating to zero the derivative on $x$ of this function
$$h_N'(x) =  x^{-2} h_N (x)\left[(N+1/4)^2 - 3x/2\right] = 0\,,$$
we find the solution  $x_*$ of this equation. It is the point of
the maximum of the function $h_N (x)$ being unique for each $N$,
$$x_* = \frac 2 {3} (N+1/4)^2\,,$$
$$h_N(x_*) = \left(\frac 3{2e}\right)^{3/2} (N + 1/4)^{-3}\,.$$

Hence, the estimation of the $N$th remainder is
$$|g(x)-g_{N-1}(x)| \le \sqrt{\frac 2 {\pi}}a_Nh_N(x_*)\ =\ \frac 32
\sqrt{\frac 3{\pi e}}
\left(\frac{(2N)!}{4^{N}(N!)^2}\right)(N+1/4)^{-2}\,.$$
Further, we estimate the coefficient $a_N$ having done more
transparent the obtained estimation. It is made by the following way
$$a_N\ =\ \frac {(2N - 1)!!}{2^N\, N!}\ =\ \prod_{l=1}^N
\left(\frac {2l - 1}{2l}\right)\ =\
\prod_{l=1}^N\left(1 - \frac {1}{2l}\right)\ =$$
$$=\
\exp\left[\sum_{l=1}^N \ln(1 - (2l)^{-1})\right]\ <
\exp\left[-\frac 12 \sum_{l=1}^N l^{-1}\right]\ <
\exp\left[-\frac 12 (1 + \ln\,N)\right]\ =$$
$$=\ \frac {e^{- 1/2}}{\sqrt{N}}\,,$$
in view of validity of inequalities  $\ln(1 - x) < -x$ at $x > 0$ and
$$\sum_{l=1}^N \frac 1l\ >\ 1 + \int\limits^N_1 \frac {dx}x = 1 + \ln\,N\,.$$
Since  $\sqrt{3/\pi} < 1$ then Eq.(9) takes place.\ $\blacksquare$

{\bf 3.}\ \ Now,\ \ we estimate\ \ the
approximation accuracy\ \ of probabilities
${\rm Pr}\{J_T[w] > c\}$ which are
obtained on the basis of functions  $g_N(\cdot)$, $N=1, 2, ...$\,.
Since $f(x) = T^{-2}g(T^{-2}x)$ then
$${\rm Pr}\{J_T [w] > c\}\ =\ 1 - T^{-2}\int\limits^c_0 g\left(T^{-2}x\right)
dx\ \equiv\ 1 - R (c)\,,$$
where
$$R_N (c)\ =\ \int\limits^{c/T^2}_0 g_{N-1}(x) dx\,.$$

Designating the righthand side of the inequality (5) by $Q_N (x)$,
we have $|g(x) - g_{N-1}(x)| \le Q_N (x)$.
Further, we determine the function
$$P_N (c)\ \equiv\ 1\ -\ R_N(c)\,,\quad R_N(c)\ =\ \int\limits^{c/T^2}_0
g_{N-1}(x) dx\,.$$

Our problem is\ \ the reception\ \ of the top estimation\ \ of the deviation
$\left| {\rm Pr}\{J_T [w] > c\} - P_N (c)\right|$. From the inequality (5),
it follows
$$- Q_N (x) \le g(x) - g_{N-1}(x) \le Q_N (x)\,.$$
Integrating between limits 0 and $c/T^2$, we obtain
$$- \int\limits_0^{c/T^2} Q_N (x) dx\le  R(c) -
\int\limits_0^{c/T^2}g_{N-1}(x) dx \le \int\limits_0^{c/T^2}Q_N (x) dx\,.$$
Hence,
$$|R(c) - R_N (c)|\ =\ \left|\int\limits^{c/T^2}_0 \left(g(x) - g_{N-1}(x)\right)
dx\right|\ \le\ \int\limits^{c/T^2}_0 Q_{N}(x) dx\,.$$
This gives the desired estimation
$$\left| {\rm Pr}\{J_T [w] > c\} - P_N (c)\right|\ =\
|R(c) - R_N (c)|\ \le\ \int\limits^{c/T^2}_0 Q_{N}(x) dx\,. \eqno (10)$$

At last, we calculate the integral in the righthand side. Due to this, the
estimation (10) becomes obvious
$$\int\limits^{c/T^2}_0 Q_{N}(x) dx\ =\ a_N  \sqrt{\frac 2\pi}
\int\limits^{c/T^2}_0 \exp \left(- \frac {(N + 1/4 )^2}x\right)\frac
{dx}{x^{3/2}}\,.$$
Replacement of the integration variable $y = x^{-1/2}$, $dy = - dx/2x^{3/2}$
results in the formula
$$\int\limits^{c/T^2}_0 Q_{N}(x) dx\ =\ \sqrt{\frac 8 \pi}
\frac {a_N}{N+1/4}
\int\limits^\infty_{\frac {T (N+1/4)}{c^{1/2}}} e^{- y^2} dy\ =\
\frac {\sqrt{2} a_N}{N + 1/4}\, {\rm Erfc} \left[\frac {T(N+1/4)}{c^{1/2}}
\right]\,.$$
From here, using Erfc$(x) \le 1$, we find the following estimation being
uniform on parameters $c$ and $T$
$$\int\limits^{c/T^2}_0 Q_{N}(x) dx\ \le\
\sqrt{\frac 2 {e N^3}}\,. $$
More exact estimation which takes into account the order of parameters $c$
and $T$ is obtained by using the standard inequality
Erfc$(x) < (\sqrt{\pi} x)^{-1} \exp (- x^2)$,
$$\int\limits^{c/T^2}_0 Q_{N}(x) dx\ \le\ \sqrt{\frac {2c}\pi}\,\frac
{a_N}{T (N+ 1/4)^2} \exp \left(- \frac {T^2(N+1/4)^2}{c}\right)\ <$$
$$<\ \sqrt{\frac {2c}{\pi e}} \left(T N^{5/2}\right)^{-1}\exp \left(
- (TN)^2/c\right)\,.$$

\end{document}